\documentclass[openacc]{rstransa}%%%%where rstrans is the template name

\usepackage{graphicx}
\graphicspath{{Images/}}
\usepackage{multirow}
\usepackage[normalem]{ulem}
\usepackage[square,sort,comma,numbers]{natbib}
\setcitestyle{square}
\usepackage{lscape}
\usepackage{changepage}
\usepackage{adjustbox}
\usepackage{totcount}
\usepackage{pdflscape}

\usepackage{comment} % hide text
\usepackage{afterpage}

\usepackage[bottom]{footmisc}
\usepackage{soul} 
\usepackage{float}
\usepackage{epstopdf, epsfig}
\usepackage{graphicx}% Include figure files
\usepackage{dcolumn}% Align table columns on decimal point
\usepackage{bm}% bold math 
\usepackage[utf8]{inputenc}
\usepackage[T1]{fontenc} 
\usepackage[usenames,table,xcdraw,svgnames,dvipsnames]{xcolor}

\usepackage{rotating}

\usepackage{ulem} 

\usepackage{etoolbox}
\makeatletter
\let\clandscape=\landscape

\patchcmd{\clandscape}{\PLS@Rotate{90}}{\PLS@Rotate{-90}}{}{}
\makeatother

\begin{document}

% For us: number of papers cited - remove before submission
\newtotcounter{citenum}
\def\oldcite{}
\let\oldcite=\bibcite
\def\bibcite{\stepcounter{citenum}\oldcite}
%%%% Article title to be placed here
\title{Friction dynamics of elasto-inertial turbulence in Taylor-Couette flow of viscoelastic fluids}

\author{%%%% Author details
M.~Moazzen$^{1}$, T.~Lacassagne$^{1}$, V.~Thomy$^{2}$ and S.~A.~Bahrani$^{1}$}

%%%%%%%%% Insert author address here
\address{$^{1}$IMT Nord Europe, Institut Mines-Télécom, Univ. Lille, Centre for Energy and Environment,
F-59000 Lille, France\\
$^{2}$Univ. Lille, CNRS, Centrale Lille, Univ. Polytechnique Hauts-de-France, UMR 8520 - IEMN – Institut d’Electronique de Microélectronique et de Nanotechnologie, F-59000 Lille, France\\
}

%%%% Subject entries to be placed here %%%%
\subject{Fluid Mechanics - Physics}

%%%% Keyword entries to be placed here %%%%

%%%% Insert corresponding author and its email address}
\corres{S. Amir Bahrani:
\email{amir.bahrani@imt-nord-europe.fr}}

%%%% Abstract text to be placed here %%%%%%%%%%%%
\begin{abstract}
Dynamic properties of elasto-inertial turbulence (EIT) are studied in a Taylor-Couette geometry. EIT is a chaotic flow state that develops upon both non-negligible inertia and viscoelasticity. A combination of direct flow visualisation and torque measurement allows to verify the earlier onset of EIT compared to purely inertial instabilities (and inertial turbulence). The scaling of the pseudo-Nusselt number with inertia and elasticity is discussed here for the first time. Variations in the friction coefficient, temporal frequency spectra, and spatial power density spectra highlight that EIT undergoes an intermediate behavior before transitioning to its fully developed chaotic state that requires both high inertia and elasticity. During this transition the contribution of secondary flows to the overall friction dynamics is limited. This is expected to be of great interest in the aim of achieving efficiency mixing at low drag and low but finite Reynolds number.

\end{abstract}
%%%%%%%%%%%%%%%%%%%%%%%%%%%

%\clearpage
%%%%%%%%%% Insert the texts which can accomdate on firstpage in the tag "fmtext" %%%%%

\begin{fmtext}
\end{fmtext}

\maketitle

\section{Introduction}\label{sec:introduction}
 
The Taylor-Couette (TC) geometry consists in two concentric cylinders, with either one or both cylinders rotating. Since the seminal work of Taylor \citep{Taylor1923VIII.Cylinders}, it has been extensively used by researchers \citep{Andereck1992OrderedFlow} thanks to its simplicity of use which allows to easily study flow instabilities. In the most familiar case where only the inner cylinder is rotating and the outer cylinder is stationary, a non-dimensional control parameter is the Reynolds number, which is defined as $\mathcal{R}=\rho\Omega r_i \delta / \mu$, where $\mu$, $\rho$, $\Omega$ are the dynamic viscosity, fluid density and the rotational speed of the inner cylinder, respectively, and $\delta = r_o - r_i$ is the gap width, with $r_i$ and $r_o$ the inner and outer radii. The geometry can be characterised by two non dimensional parameters: the aspect ratio $\Gamma = h / \delta$ and radius ratio $\eta = r_{i} / r_{o}$, where $h$ is length of the cylinder. Alternatively, the curvature ratio $\kappa=r_i/\delta$ can be used. The aforementioned parameters are known to have an influence on stability, not only in Newtonian but also in non-Newtonian fluids.
Indeed, the TC geometry is also widely used in the study of complex fluids \citep{Topayev2019,Fardin2014}, such as dilute polymer solutions \citep{Song2021DirectFlow,Liu2013ElasticallyInsight} or suspensions of particles \citep{Moazzen2022TorqueSuspensions} among others. In particular, polymeric liquids flow which exhibit viscoelastic behavior have been studied with great interest, due to the existence of sets of specific flow regimes, and motivated by the ubiquity of viscoelasticity in daily life, industrial and natural applications, such as biology, pharmaceutics, paints, among others \cite{Schrimpf2021}. 

The mechanism of instability in these fluids are different from those that occur in Newtonian cases. In Newtonian fluids, the instability comes from the destabilizing effect of the centrifugal force gradient (which comes from variations of kinetic momentum), and overcoming of it on the stabilizing effect of viscous drag force. In such fluids, at low Reynolds number, a purely azimuthal uniform shear flow develops, which is called circular Couette flow (CCF). It eventually becomes unstable upon increasing $\mathcal{R}$ as explained above, and secondary flows appear as axisymmetric counter-rotating vortices called Taylor vortex flow (TVF). Further increase of Reynolds number creates non-axisymmetric sinusoidal axial oscillations called wavy Taylor vortex flow (WVF). Eventually, additional wavelentghs appear and the flow transitions to turbulence \citep{Grossmann2016,Huisman2014MultipleFlow,Tuckerman2020,Takeda1999,Dutcher2009Spatio-temporalFlows}. 

In non-Newtonian, viscoelastic fluids, the mechanism of instability, and subsequently its flow transition, is different. Polymers solutions are common viscoelastic fluids. Polymers are high molecular weight molecules made of a large number of monomers connected with covalent bonds, resulting in long linear, branched or network chains \citep{Coussot2014RheophysiqueETATS}). The arrangement (conformation) of the polymer chain at the rest condition is in the way that have maximum conformational entropy \citep{Gennes1979ScalingPhysics}. When the polymer coil is stretched, e.g  because of an applied stress or deformation, it tends to recover its lost maximum entropy energy and return to its equilibrium chain structure. Due to this entropic tendency of polymers, elastic stresses are created in the chain which as a result of the stress difference between the flow direction and the direction perpendicular to it (direction of shear), which doesn't exist in Newtonian fluids. In rotational flow such as Taylor-Couette Flows (TCF), curved streamlines induce a hoop stress, balanced by an adverse pressure gradient in the radial direction. A flow perturbation may cause a fluid particle to move towards a region of enhanced stretching, enhancing the local hoop stress and destabilizing the flow \cite{Burghelea2020TransportFluids}. Moreover, a part of the chain's elastic deformation energy can be released elsewhere in the flow, further promoting instability. The elastic behavior of polymer solutions thus highly depend on deformation rates, and relaxation time of polymer chains. The degree of elastic response of a fluid subjected to a shear rate $\dot{\gamma}$ is quantified by the Weissenberg number $\mathcal{W}i$, defined in the case of TCF as $\mathcal{W}i=\lambda_e \dot{\gamma}$ with $\dot{\gamma}=\Omega r_i/ \delta$ the nominal shear rate in the gap. The elastic number El is then defined by
$\mathrm{El}=\frac{\mathcal{W}i}{\mathcal{R}}=\frac{\lambda_e }{\lambda_v}=\frac{\lambda_e \mu}{\rho \delta^2}$
and represents the competition between inertial and elastic effects, with $\lambda_v=\rho \delta^2/\mu$ the viscous characteristic time. The resulting El depends only on the geometrical parameters and the properties of the fluid (which may themselves be shear-rate dependent, see below). El allows to classify fluids into 3 groups: weak ($\mathrm{El}<10^{-2}$), moderate ($10^{-2}<\mathrm{El}<1$) and strong elasticity ($\mathrm{El}>1$) \citep{Dutcher2011EffectsFlows,Dutcher2013EffectsFlows,Latrache2016Defect-mediatedFlow}. Based on the elasticity level, various instability and transition scenarios are observed. In the range of very low elasticity (i.e, $\mathrm{El}\ll 1$) the elastic effects are very weak compared to inertia effects and observed flow transition are comparable to the Newtonian case (CCF$\rightarrow$TVF$\rightarrow$WVF) as $\mathcal{R}$ increases \citep{Crumeyrolle2002ExperimentalSolutions,Dutcher2011EffectsFlows} with slightly shifted critical conditions $\mathcal{R}_c$ because of presence of light amount of polymer.

At high values of elasticity, in the case of vanishing $\mathcal{R}$, a purely elastic CCF-TVF transition is observed \citep{Larson1990AFlow}. Subsequently in the case of elevated $\mathcal{W}i$ another transition will lead  to a chaotic regime called elastic turbulence \citep{Steinberg2019ScalingTurbulence,Groisman2004ElasticSolutions,Larson2000TurbulenceInertia}, which exhibits turbulent like-characteristics in absence of inertia. When neither $\mathcal{R}$ nor $\mathrm{El}$ can be neglected, we find ourselves in the domain of elasto-inertial transitions. In particular, primary and secondary elasto-inertial instabilities manifest themselves in non-axisymmetric flow states \citep{Baumert1999AxisymmetricFlow,Latrache2016Defect-mediatedFlow,Mohammadigoushki2017Inertio-elasticSystem,Groisman1996Couette-TaylorSolution,Dutcher2013EffectsFlows,Lacassagne2020VortexFlow}. An increase in inertia ($\mathcal{R}$) or elasticity (El) leads these pre-chaotic behaviors to transition to strongly unsteady states: "disordered oscillations" (DO) \citep{Groisman1996Couette-TaylorSolution}, "defect mediated turbulence (DMT) \cite{Latrache2016Defect-mediatedFlow}, "spatio-temporal intermittency" (STI) \citep{Latrache2012TransitionCell} or "merge-split transitions" (MST) \citep{Lacassagne2020VortexFlow}, and all contribute to a gradual transition to elasto-inertial turbulence (EIT) \citep{Dutcher2013EffectsFlows,Liu2013ElasticallyInsight,Lacassagne2020VortexFlow}. A summary of several flow transition observed experimentally, as a function of geometrical parameter and viscoelastic fluid properties, are listed in table \ref{tab:geometrical_parameter}.

The possibility of triggering such chaotic behavior opens extremely interesting perspectives in terms of mixing and intensification of transfers at low $\mathcal{R}$. While transition scenarios are now relatively well identified in the literature, several questions remain to be tackled: what are the characteristics of EIT in TCF? What is the dynamic behavior of these flows in terms of friction and energy dissipation ? This work aims at addressing this last point in particular, by reporting for the first time friction and spatio-temporal properties of TCF of constant viscosity polymer solutions with shear-dependent viscoelasticity.

\afterpage{
% \usepackage{lscape}
\begin{landscape}
\begin{table}[ht!]
\centering
\begin{tabular}{lcccccccccc}
\rowcolor[HTML]{222222} 
{\color[HTML]{FFFFFF} Study} & {\color[HTML]{FFFFFF} $\eta$} & {\color[HTML]{FFFFFF} $\Gamma$} & {\color[HTML]{FFFFFF} $\delta$ (mm)} & {\color[HTML]{FFFFFF} $\kappa$} & {\color[HTML]{FFFFFF} solution} & {\color[HTML]{FFFFFF} $\mu_{p} / \mu $} & {\color[HTML]{FFFFFF} El} & {\color[HTML]{FFFFFF} $\mathcal{R}_{max}$} & {\color[HTML]{FFFFFF} $d \mathcal{R}/dt^{*}$} & {\color[HTML]{FFFFFF} FTO} \\
Present study & 0.914 & 10 & 1.5 & 0.093 & HPAM/W & 0.075 & 0.05-0.1 & 200 & <0.01 & CCF-SVF-RSW-MWVF \\
 & \multicolumn{1}{l}{} &  & \multicolumn{1}{l}{} & \multicolumn{1}{l}{} & /G/S & 0.086-0.221 & 0.11-0.35 & \multicolumn{1}{l}{} &  & CCF-EIT \\
\rowcolor[HTML]{D9D9D9} 
Lacassagne et. & 0.77 & 21.56 & 6.26 & 0.289 & PAAm/W/G & 0.148 & 0.2263 & 200 & 0.33 & CCF-TVF-RSW-EIT \\
\rowcolor[HTML]{D9D9D9} 
al., (2020) \cite{Lacassagne2020VortexFlow} &  &  &  &  &  &  &  &  &  &  \\
Martinez-Arias $\&$ & 0.909 & 30 & 5 & 0.1 & PEO/PEG & 0.083-0.325 & 0.06-0.17 & 160 & <0.6 & CCF-TVF-RSW-EIT \\
 Peixinho (2017)\cite{Martinez-Arias2017TorqueSolutions} &  &  & \multicolumn{1}{l}{} &  & /W/IPA & 0,78 & 0.71-1.09 & 84 &  & CCF-RSW--EIT \\
\rowcolor[HTML]{D9D9D9} 
Dutcher et. & 0.912 & 60.7 & 6.69 & 0.096 & PEO/W/G & 2.82 & 0.1-0.21 & 200-250 & 0.68 & CCF-TVF-RSW-EIT \\
\rowcolor[HTML]{D9D9D9} 
al., (2013) \cite{Dutcher2013EffectsFlows} &  &  &  &  &  &  &  &  &  &  \\
Dutcher et. & 0.912 & 60.7 & \multicolumn{1}{l}{6.69} & 0.096 & PEO/W/G & 0.3 & 0.00047 & 200-250 & <0.68 & CCF-TVF-WVF-...-TTV \\
al., (2011) \cite{Dutcher2011EffectsFlows} &  &  & \multicolumn{1}{l}{} & \multicolumn{1}{l}{} & \multicolumn{1}{l}{} & 0.93 & 0.0017 &  &  & CCF-TVF-WVF-MWVF-WVF-MWF-CWV-WTV-MT \\
 &  &  & \multicolumn{1}{l}{} & \multicolumn{1}{l}{} & \multicolumn{1}{l}{} & 0.92 & 0.0054 &  &  & CCF-TVF-WVF-MWVF-WVF-WTV / CWV \\
 &  &  & \multicolumn{1}{l}{} & \multicolumn{1}{l}{} & \multicolumn{1}{l}{} & 0.78 & 0.023 &  &  & CCF-TVF-WVF-MWVF-WVF \\
\rowcolor[HTML]{D9D9D9} 
Crumeyrolle et. & 0.883 & 47 & 5.9 & 0.132 & PEO/W & 0 - 3.45 & 0.002-0.03 & 200 & N/A & CCF-TVF-WVF \\
\rowcolor[HTML]{D9D9D9} 
al., (2005) \cite{Crumeyrolle2005InstabilitiesSystem} &  &  &  &  & \multicolumn{1}{l}{\cellcolor[HTML]{D9D9D9}} & 5.32-12.4 & 0.07-0.5 &  &  & CCF-RSW \\
Groisman et. & 0.829 & 74 & \multicolumn{1}{l}{7} & \multicolumn{1}{l}{0.2} & PAAm/W & 0.82 & 0.025 & N/A & N/A & CCF-TVF-WVF \\
al., (1998) \cite{Groisman1998ElasticFlow} &  &  & \multicolumn{1}{l}{} & \multicolumn{1}{l}{} & \multicolumn{1}{l}{/saccharose} &  & 0.03-0.08 & \multicolumn{1}{l}{} & \multicolumn{1}{l}{} & CCF-TVF-RSW-DO \\
 &  &  & \multicolumn{1}{l}{} & \multicolumn{1}{l}{} & \multicolumn{1}{l}{} &  & 0.09-0.15 & \multicolumn{1}{l}{} & \multicolumn{1}{l}{} & CCF-DO \\
 &  &  & \multicolumn{1}{l}{} & \multicolumn{1}{l}{} & \multicolumn{1}{l}{} &  & 0.2-27 & \multicolumn{1}{l}{} & \multicolumn{1}{l}{} & CCF-DO \\
\rowcolor[HTML]{D9D9D9} 
Groisman et. & 0.708 & 54 & 7.85 & 0.413 & PAAm/W & 0.008-0.25 & 0.1-0.15 & N/A & N/A & CCF-TVF-RSW-DO \\
\rowcolor[HTML]{D9D9D9} 
al., (1996) \cite{Groisman1996Couette-TaylorSolution} &  &  &  &  & /saccharose &  & 0.15-0.22 &  &  & CCF-TVF-DO \\
\rowcolor[HTML]{D9D9D9} 
 &  &  &  &  &  &  & 0.22-0.34 &  &  & CCF-DO \\
Groisman et. & 0.708 & 54 & 7.85 & 0.413 & \multicolumn{1}{l}{PAAm/W} & 0.78 & 0.023-0.033 & N/A & N/A & CCF-RSW-DO \\
al., (1993) \cite{Groisman1993ExperimentsSolutions} &  &  & \multicolumn{1}{l}{} & \multicolumn{1}{l}{} & \multicolumn{1}{l}{/saccharose} &  &  & \multicolumn{1}{l}{} & \multicolumn{1}{l}{} & 
\end{tabular}
\caption{Some experimentally observed flow transition patterns in viscoelastic Boger fluids (constant viscosity $\mu$ assumed) with different fluid properties and geometrical parameters (rotating inner cylinder and stationary outer cylinder). DO = Disordered Oscillations, FP = Flame Pattern, SVF = Spiral Vortex Flow, MWVF = Modulated Wavy Vortex Flow, MT = Modulated Turbulence, RSW = Rotating Spiral Waves, TTV = Turbulent Taylor Vortices, CWV = Chaotic Wavy Vortex Flow. $\mu_p=\mu-\mu_s$ is the polymer contribution to the total viscosity. $\kappa=\delta/r_i$ is the curvature ratio. W = Water, G = Glycerol, PEO = Polyethylène Oxyde, PAAm = Polyacrylamide, PEG= Polyethylène Glycol, IPA = Isopropyl Alcohol. N/A = not available.}
\label{tab:geometrical_parameter}
\end{table}
\end{landscape}
 }

\section{Materials and methods}
Experiments were performed using aqueous Boger solutions of high molecular weight polymer of partially hydrolysed polyacrylamide (HPAM, $M_w = 15 - 20 \times 10^6$ g/mol). At first, a stock aqueous polymer solution of 1000 ppm was prepared. Samples from this solution were then dissolved in pure water and mixed in glycerol in order to obtain different concentrations $\tilde{c}_p$ of 25, 50, 100, 150, 200, 250, 300, 350 ppm with base solution similar to that of our previous study \citep{Moazzen2022TorqueSuspensions}: 41.8 $\%$ glycerol and 58.2$\%$ water (in volume) and 12.7$\%$ of salt (in mass). After preparation, the aqueous solutions are left at rest for 24 h before performing any other manipulation.

\begin{figure}[ht!]
    \centering
    \includegraphics[width=1\columnwidth]{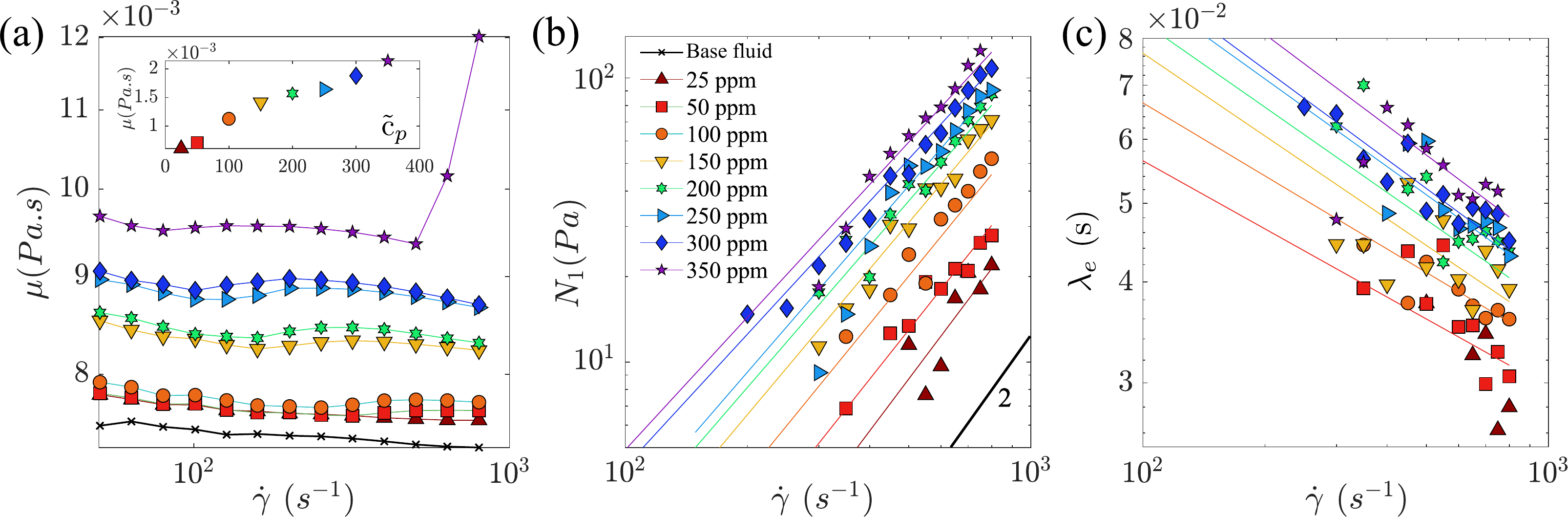}
    \caption{\label{rheologie_ensemble} (a) Shear viscosity as a function of shear rate for HPAM polymer solution at various concentrations. The observed jump in viscosity at a concentration of 350 ppm (after 600 s$¨{-1}$) is related to an elastic instability in the rheometer. (b) Measured first normal stress N1 of the solutions. A line with slope of 2 is plotted as a guide to the eye. (c) Relaxation time obtained using our experimental methodology. Colour lines are fits to the experimental data.}
\end{figure} 
\raggedbottom

Rheological behavior of all working fluids was characterized using a rotational rheometer (Anton-Paar MCR 302) equipped with a cone-plane geometry $\left(50 \mathrm{~mm} / 1^{\circ}\right)$ with truncation gap of $\theta=$ $0.104 \mathrm{~mm}$ at a constant temperature of $22{ }^{\circ} \mathrm{C}$. In order to find the viscosity of samples, steady-shear viscosity measurements were performed on a shear-rate interval of \mbox{0.01/s $< \dot{\gamma} <$ 800/s}. Samples viscosity remained constant over this range of shear rates. The flow curves for all samples are reported in figure \ref{rheologie_ensemble}. In order to evaluate the dynamic, shear rate dependent, relaxation times $\lambda_e $, the protocol detailed by \cite{Bahrani2022PropagationModel}, based on normal force measurement, was used. It consisted of several incremental steps during which constant shear rate was imposed. Subtracting the means of two values of the first normal stress difference $N_{1}$ for each step gives a way to resolve the instrumental drift of the normal force and correct $N_{1}(\dot{\gamma})$ values. A second correction is performed to remove the contribution of fluid inertia to the normal force given by the rheometer $N_{1,tot}$, such that $N_{1} = N_{1,tot} + 0.15 \rho \Omega^{2} R^{2}$ with $\Omega =\tan \left(\theta\right) \dot{\gamma}$ the angular velocity (the 0.15 prefactor corresponds to inertial and secondary flow corrections and was proposed by \cite{ChristopherW.Macosko1994}). $N_1$ can be expressed as a power-law function of the shear rate; $N_{1} = \Psi \dot{\gamma}^\epsilon$, where $\Psi$ and $\epsilon$ are constants. $\epsilon=2$ implies that the behavior follows the Oldroyd-B model \citep{1950OnState}: $N_{1}=2 (\mu-\mu_s) \lambda_e \dot{\gamma}^2$, and the viscosity $\mu$ is dominated by the Newtonian solvent contribution, $\mu_{s}$. However, unlike the Oldroyd-B model, the relaxation time here also follows a power law (shear-rate dependent) function as $\lambda_e = a \dot{\gamma}^{b}$, where $a$ and $b$ are constants. As polymer concentration increases, $b$ become more negative which means that $\lambda_e$ becomes more sensitive to shear rate. A summary of the aforementioned coefficients is shown in table \ref{coefficient}. This advanced viscoelasticity characterisation protocol allows to account for the effective shear-dependency of $\lambda_e$, and thus El, in constant viscosity fluids and thus increase the accuracy on the critical El values detection. As expected from \cite{Casanellas2016TheMicroflows} the method performs better for higher polymer concentrations, with less noise on the $N_1$ and $\lambda_e$ data. It here results in a poor fitting of $\lambda_e$ data for the 25 ppm case only.

\begin{table}[ht!]
\centering
\begin{tabular}{cccccccccccc}
 &  & \multicolumn{1}{l}{Coef.} &  & \multicolumn{8}{c}{$\tilde{\mathrm{c}}_p$ (ppm)} \\
 \cline{3-3} \cline{5-12} 
 &  &  &  & 25 & 50 & 100 & 150 & 200 & 250 & 300 & 350 \\
\multicolumn{1}{l}{} & \multicolumn{1}{l}{} & \multicolumn{1}{l}{} & \multicolumn{1}{l}{} & \multicolumn{1}{l}{} & \multicolumn{1}{l}{} & \multicolumn{1}{l}{} & \multicolumn{1}{l}{} & \multicolumn{1}{l}{} & \multicolumn{1}{l}{} & \multicolumn{1}{l}{} & \multicolumn{1}{l}{} \\
\multirow{2}{*}{$\lambda_e$} &  & a &  & 0.057 & 0.205 & 0.265 & 0.367 & 0.42 & 0.48 & 0.52 & 0.59 \\
 &  & b &  & -0.1 & -0.28 & -0.3 & -0.34 & -0.35 & -0.36 & -0.37 & -0.375 \\
\multirow{2}{*}{$N_1$} &  & $\Psi$ &  & $6.5e^{-5}$ & $1.8e^{-4}$ & $3.8e^{-4}$ & $8e^{-4}$ &$1.3e^{-3}$ & $1.5e^{-4}$ & $3.4e^{-3}$ & $3.9e^{-3}$ \\
 &  & $\epsilon$ &  & 1.9 & 1.8 & 1.75 & 1.7 & 1.65 & 1.65 & 1.55 & 1.55\\
 \cline{3-12}
\end{tabular}
\caption{Rheological parameter of HPAM polymer solution derived by fitting an Oldroyd-B model in order to find relaxation time, $\lambda_e$. The solutions follow the relation of $\lambda_e = a\dot{\gamma}^b$ and $N_{1}=\Psi \dot{\gamma}^{\epsilon}$, where b and $\epsilon$ govern the viscoelastic behavior.}
\label{coefficient}
\end{table}

\begin{figure}[ht!]
    \centering
    \includegraphics[width=0.8\columnwidth]{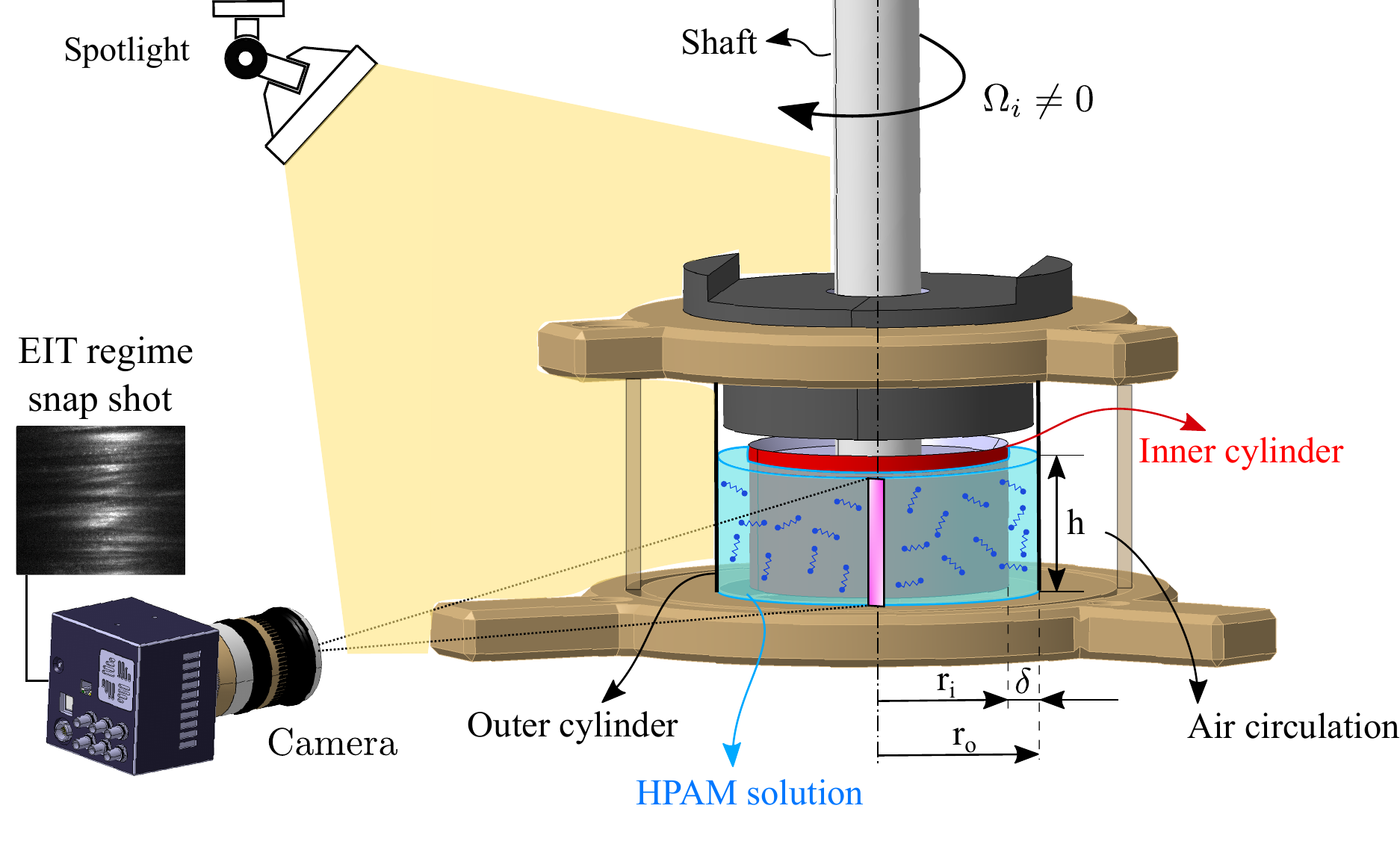}
  \caption{Schematic of the Taylor-Couette apparatus  mounted on the Anton-Paar MCR-102 rheometer with camera and light position for visualisation. As shown in the figure, the upper end and lower end of the gap are a free surface and a stationary wall, respectively.} \label{setup} 
\end{figure}

The TCF experiments were performed in a Taylor-Couette cell mounted on the same rheometer as illustrated in figure \ref{setup}. The geometrical parameters were: $\delta = 1.5$ mm, $\eta = 0.914$ and $\Gamma =  10$. In the present study, only ramp-up experiments were performed (slow acceleration of the inner cylinder), combining torque measurements and visualisation using Iriodin particles ($\sim 0.1 \%$ in mass) and a light source, following a protocol detailed in our previous work  \cite{Moazzen2022TorqueSuspensions}. The inner cylinder acceleration rate was $0.0082< \frac{d \mathcal{R}}{d t^{*}}=\frac{\rho^{2} r_ {i} \delta^{3}}{\mu^{2}} \frac{d \Omega}{d t}  < 0.01$, and the temperature was 22 ($\pm$ $0.4$) °C.

\section{Results and discussion}

\subsection{Torque measurements}

\begin{figure}
    \centering
    \includegraphics[width=1\columnwidth]{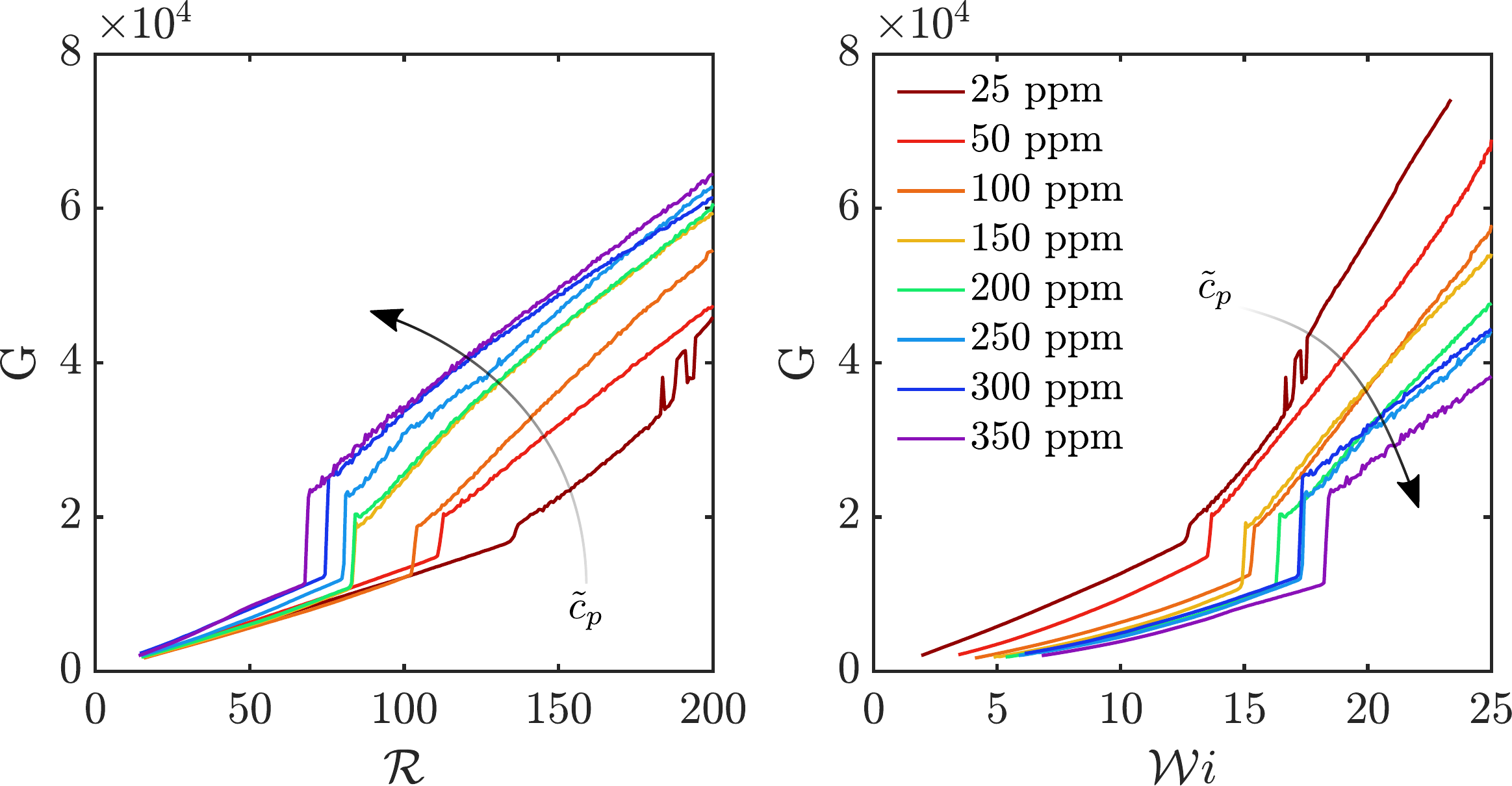}
    \caption{Plots of $G$ as a function of $\mathcal{R}$ (left) and $\mathcal{W}i$ (right) for all polymer concentrations $\tilde{c}_p$.}
    \label{fig:Gplots}
\end{figure}

Let $\mathcal{T}$ be the raw measured torque on the rheometer shaft and $G= \mathcal{T}/\mathcal{T}_v=\mathcal{T} / 2 \pi h\left(\mu^{2} / \rho \right)$ the dimensionless torque. G is thus simply a non-dimensional version $\mathcal{T}$, scaled by torque-scale $\mathcal{T}_v$.  Figure \ref{fig:Gplots} shows plots for $G$ as a function of $\mathcal{R}$ (left) and $\mathcal{W}i$ (right) for all polymer solutions. The discontinuity in $G$ values indicates the onset of a secondary flow, as will be discussed in section \ref{sec:transition}. It appears that $G$ increases with both increasing $\mathcal{R}$ and increasing $\mathcal{W}i$, since the shear-rate increases. However, the increase rate of $G$ after the discontinuity is reduced as $\tilde{c}_p$ increases, which is a key feature of the drag dynamics of the unsteady flow state, as will be detailed in sections \ref{sec:scaling} and \ref{sec:drag}.

\subsection{Transitions, flow states, and critical $\mathcal{R}$}
\label{sec:transition}

\afterpage{
\begin{landscape}
\begin{figure}[ht!]
   \hspace{-2cm}
%    \centering
    \includegraphics[width=1.2\columnwidth]{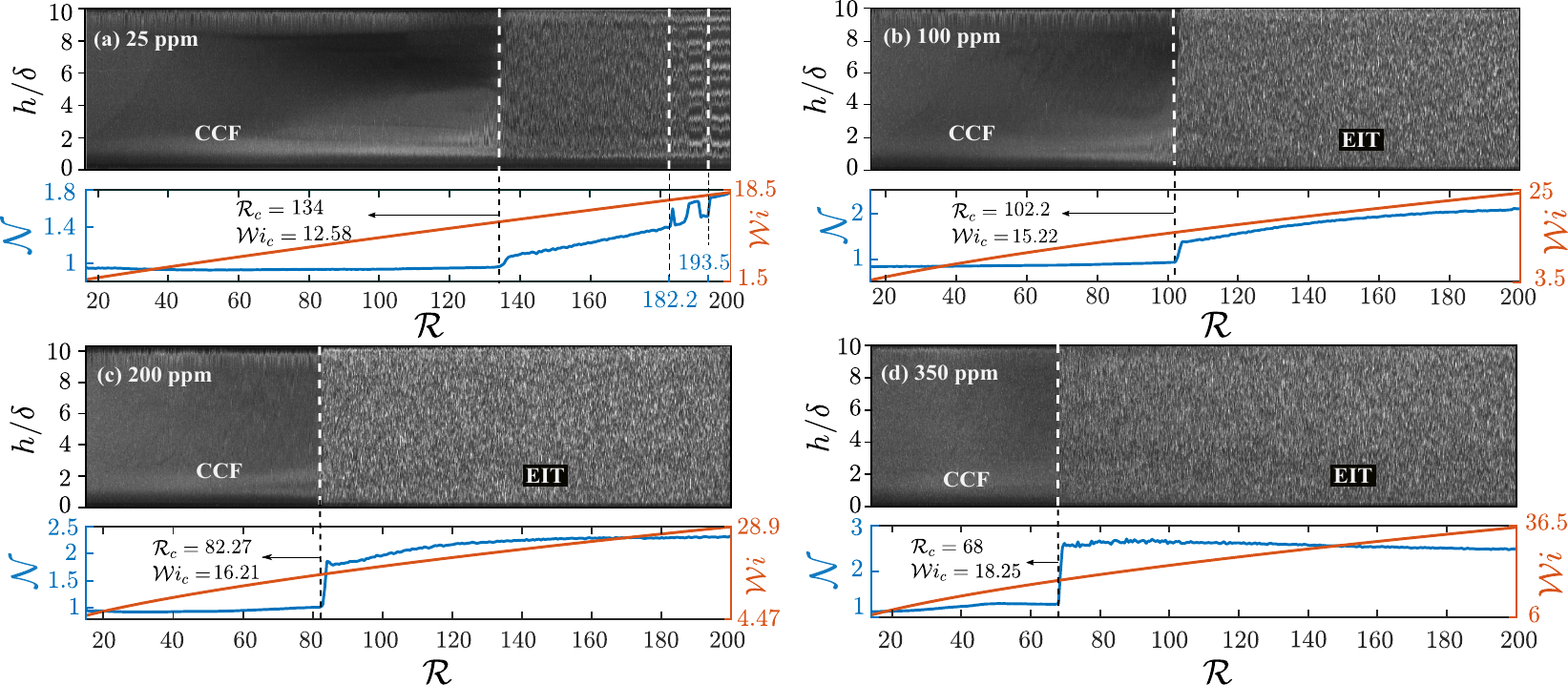}
    \caption{Space-$\mathcal{R}$ diagram (top), Nusselt-$\mathcal{R}$ (bottom plots, left axis) and $\mathcal{W}i$-$\mathcal{R}$ (bottom plots, right axis) plots. The plots show the transitions from CCF to RSW (Rotating Spiral Waves) and TVF for a concentration of (a) 25 ppm and from CCF to EIT for (b) 100 ppm (c) 200 ppm (d) 350 ppm.} \label{Nu_st}
\end{figure}
\end{landscape}
}

Flow maps diagram (space-$\mathcal{R}$ diagrams), for $\tilde{c}_p$= 25, 100, 200, 350 ppm, coupled with plots of $\mathcal{W}$i and of the effective (pseudo) Nusselt number, $\mathcal{N}$ as a function $\mathcal{R}$ is presented in Figure \ref{Nu_st}. $\mathcal{W}i$ increases with yet a decreasing slope (all the more decreasing that $\tilde{c}_p$ increases), due to the shear-rate dependency of $\lambda_e $. $\mathcal{N}$ is defined as $\mathcal{N}=\mathcal{T}/\mathcal{T}_{lam}=G/G_{lam}$, with $G_{lam}=2 \eta \mathcal{R} / (1+\eta)(1-\eta)^{2}$ the dimensionless torque for the laminar flow between infinitely long cylinders. $\mathcal{N}$ allows to go further into the interpretation, compared to $G$, by scaling the (non-dimensional) torque by its laminar value, which accounts for the geometry. It represents the dissipation rate of kinetic energy \citep{Eckhardt2000ScalingFlow} and the ability of the flow to convey momentum radially. Critical values for all numbers (subscript $._c$) are found combining flow visualisation and torque measurement (jump in $\mathcal{N}$ values \citep{Martinez-Arias2017TorqueSolutions}). Here, a CCF-EIT transition is observed for $\tilde{c}_p \geq 50$ ppm (corresponding $\text{El}_c^{50} = 0.13$), as can be seen clearly in both flow map (changing alignment of Iriodin flakes from purely azimuthal to random) and $\mathcal{N}-\mathcal{R}$ diagram. This direct transition was previously observed by \cite{Groisman1996Couette-TaylorSolution}. $\mathcal{N}$ related to CCF ($\mathcal{R} < \mathcal{R}_{c}$) is almost constant and is around one. This shows that the CCF torque depends linearly on viscosity (and viscosity does not depend on the shear rate variation). The abrupt change in $\mathcal{N}$ values clearly indicates the beginning of the EIT regime.

\begin{figure}[ht!]
    \centering
    \includegraphics[width=0.6\columnwidth]{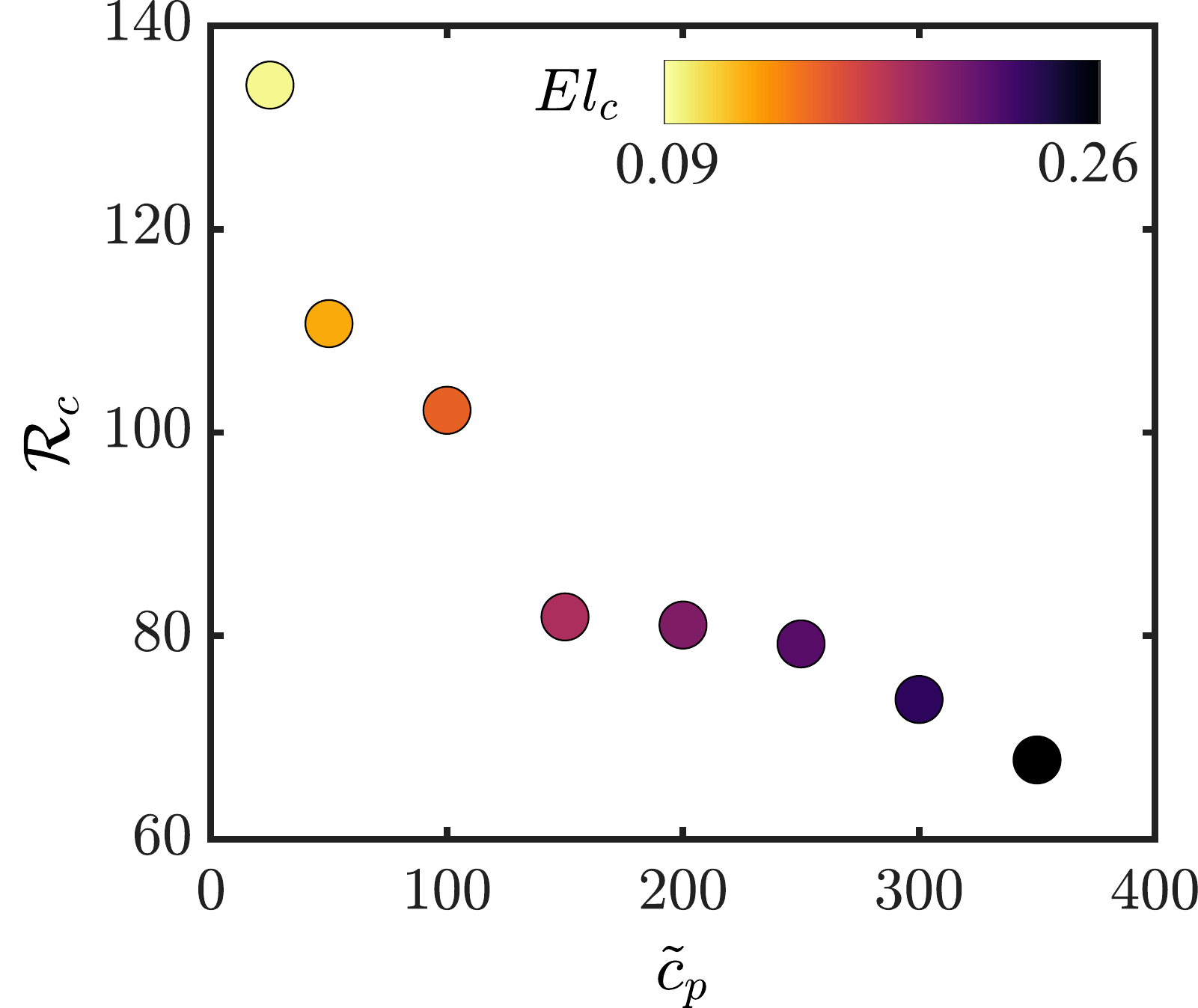}
    \caption{$\mathcal{R}_c$ for the onset of EIT (or for the primary instability, in the 25 ppm case) as a function of the $\tilde{c}_p$ with marker colours indicating $\text{El}_c$ values at the onset of EIT \label{Re_c}  }
\end{figure}

Figure \ref{Re_c} shows the $\mathcal{R}_c$ at which the CCF-EIT transition (or the primary transition, for the 25 ppm case) occurs as a function of $\tilde{c}_p$. The side color bar indicates the corresponding $\text{El}_c$. $\mathcal{R}_c$ for the 25 ppm case is 134 or $\mathcal{R}_{c}^{25} = 0.88 \mathcal{R}_{c}^{0}$, demonstrating that the CCF flow is destabilized compared to the Newtonian case (base solvent of this polymer for which the transition occurs at $\mathcal{R}_c^0$ = 150.8 in this setup \cite{Moazzen2022TorqueSuspensions}). $\mathcal{R}_{c}$ decreases as the polymer concentration or the elastic number increases. Up to 150 ppm or El = 0.18, this reduction is very strong while for $\tilde{c}_p > 150$ ppm or El > 0.18, it becomes milder, which suggests that the presence effect of polymer after this concentration is less effective. 
The lower value of $\text{El}_c$ for direct CCF-EIT ($\text{El}_c^{50}=0.12$) is here slightly lower than the value of 0.22 reported by  \cite{Groisman1996Couette-TaylorSolution}. This minor discrepancy can be ascribed to the different relaxation time estimation protocol used (ours accounting for shear-rate dependency) and to the variations in geometrical parameters \citep{Song2019TheDependence,McKinley1996RheologicalInstabilities}.

\raggedbottom

\subsection{Torque scaling in EIT}
\label{sec:scaling}

\begin{figure}[ht!]
    \centering
    \includegraphics[width=1\columnwidth]{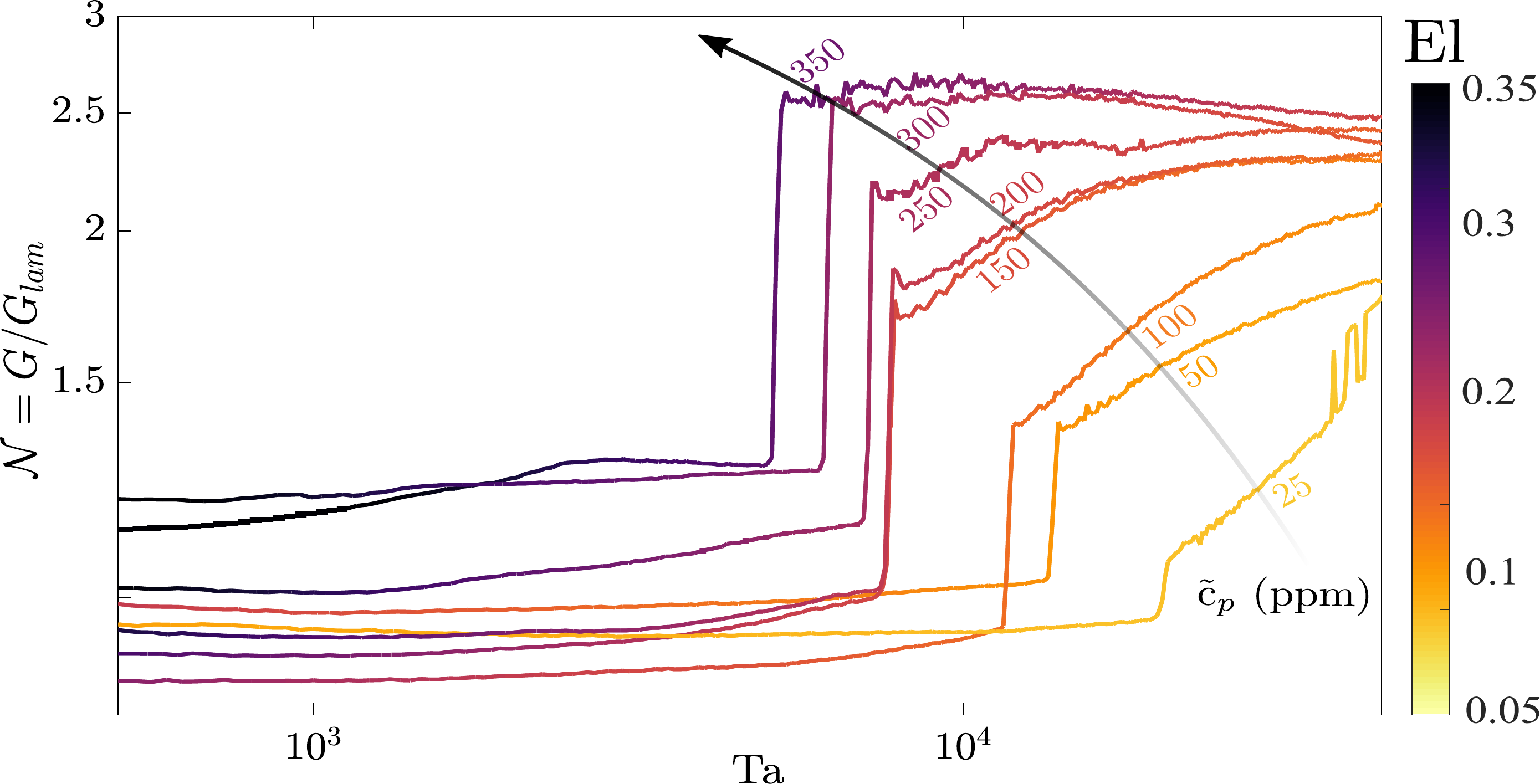}
    \caption{\label{alpha_beta_Nu_Ta} $\mathcal{N} - \text{Ta} $ curve for all concentrations with El color bar that illustrates the evolution of the elastic number during the change of Ta.}
\end{figure}

Figure \ref{alpha_beta_Nu_Ta} shows all curves for $\mathcal{N} - \text{Ta}$ at all polymer concentrations, essentially revisiting the raw data from figure \ref{fig:Gplots} but this time normalizing by laminar flow behaviour as allowed by the use of $\mathcal{N}$. $\text{Ta} = \frac{(1+\eta)^{6}}{(64\eta^{4})}\mathcal{R}^2$ is proportional to $\mathcal{R}^2$ by a geometric constant. Both numbers can be equally used to quantify flow inertia when the geometry is kept constant, but Ta is more frequently encountered in the literature for Taylor-Couette Nusselt scalings. The color map indicates the $\mathcal{R}$-dependent El value. At low Reynolds number, in the CCF state, all $\mathcal{N}$ are approximately constant, as expected. The slight variation in Nusselt number values (around 1) is due to wall effects that introduce an additional torque \citep{MartinezArias2014}. Increasing $\mathcal{R}$, up to the instability limit, there is a slight increase in the slope of $\mathcal{N}$, as noted previously by \cite{Martinez-Arias2017TorqueSolutions} or  \cite{Yi1997ExperimentalSolutions}. Passing through the critical point, an abrupt change in the value of $\mathcal{N}$ occurs. This jump intensifies with increasing $\tilde{c}_p$. Increasing the $\tilde{c}_p$, the overall value of the Nusselt number increases.

Interestingly, after the onset of EIT and as $\tilde{c}_p$ increases, the global slope of the $\mathcal{N}$-Ta curve in the EIT regime gradually decreases, evolving to a Ta-independent Nusselt number region. This evolution occurs faster as the concentration of polymer increases: the rate of change and slopes are concentration dependent, but the asymptotic behaviour appears not to be.

This can be interpreted as follows. After the onset of EIT, secondary chaotic flows arise and generate friction at the walls leading to a global increase in $\mathcal{N}$. Increasing Ta or $\mathcal{R}$, kinetic energy is injected in the flow. It is either dissipated by wall friction, which translates into an increase in $\mathcal{N}$, or by elastic dissipation by the polymer chains, which is expected not to depend on $\mathcal{R}$. Increasing $\tilde{c}_p$ comes to promoting the second mechanism over the first, reduce the share of kinetic energy dissipated by viscosity, and thus the increase in $\mathcal{N}$. This last point can be examined from another angle: by qualitatively observing the elastic threshold below which the transition to the asymptotic behaviour is gradual (and not sharp), it can be infered that even after the onset of EIT, the flow still requires a given amount of inertia and/or elasticity, i.e. a given $\mathcal{W}i$ increase, for elastic energy transfers to balance inertial ones.
 
\raggedbottom

\subsection{Elasto-inertial drag coefficient}
\label{sec:drag}

\begin{figure}[ht!]
    \centering
    \includegraphics[width=0.9\columnwidth]{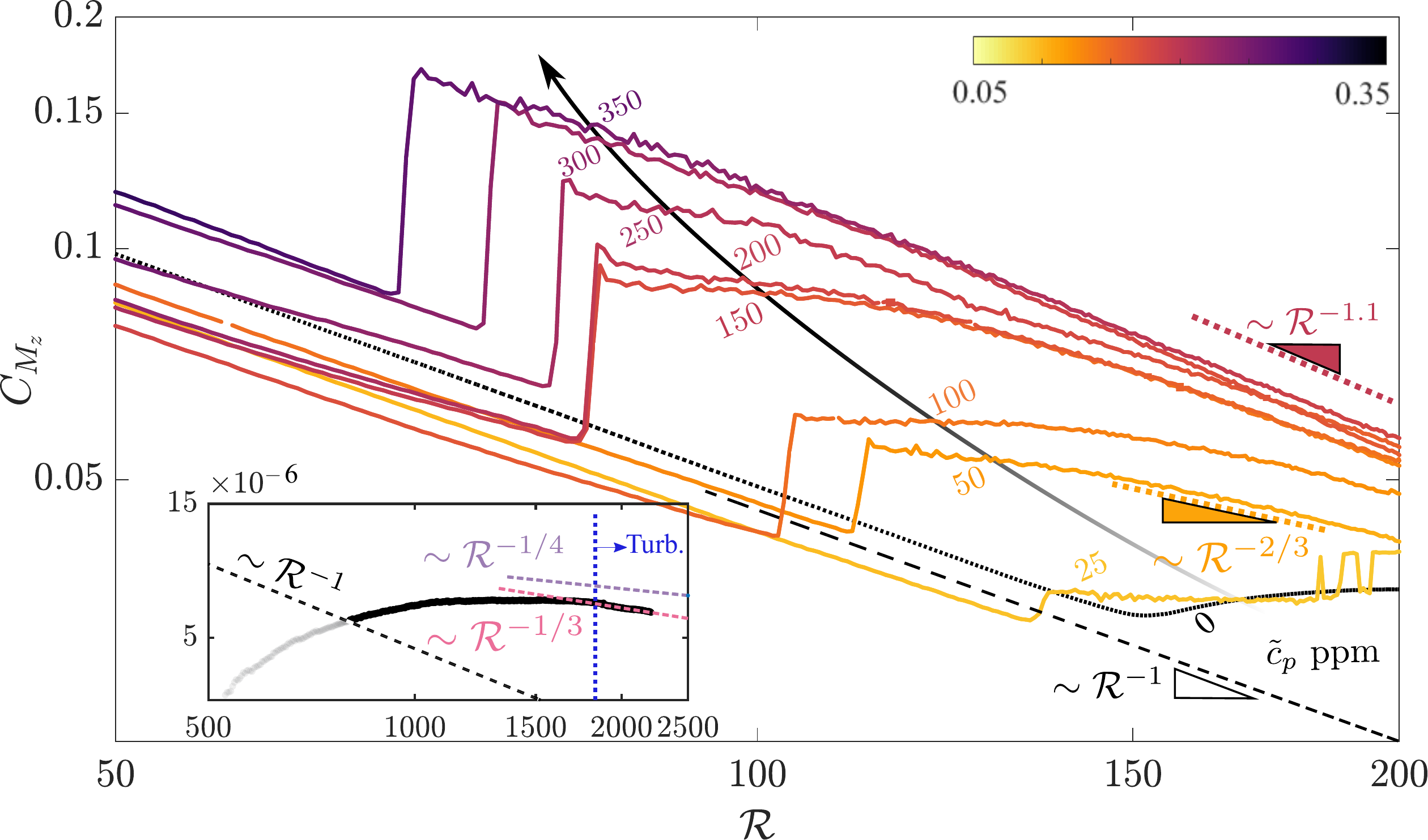}
    \caption{Friction coefficient $C_{M_z}$ as a function of $\mathcal{R}$ for all polymer concentrations (curve labels). The color scale represents El values and the dashed line the -1 slope characteristic of a laminar flow. Slopes for $\mathcal{R}^{-2/3}$ and $\mathcal{R}^{-1}$ are illustrated by dotted lines and triangle, orange and red, respectively. In inset, an experiment for pure water in turbulent flow state (full black line) is reported along with its $\sim \mathcal{R}^{-1/3}$ asymptotic trend and compared with the $\sim \mathcal{R}^{-1/4}$ trend of \cite{Greidanus2015TurbulentRotation}. The gray part of the curves correspond to points below the rheometer's minimum torque capacity and should be discarded.\label{c_m}  }
\end{figure}

In order to compare the data with Newtonian (laminar or turbulent)  experiments and references \cite{Greidanus2015TurbulentRotation}, curves from figure \ref{alpha_beta_Nu_Ta} can be re-scaled to display the friction coefficient $C_{M_z}$ defined as $C_{M_z} \sim \mathcal{N}/\mathcal{R}$ \cite{Moazzen2022TorqueSuspensions}, shown in figure \ref{c_m}.

In the Newtonian case, the onset of TVF is known to stop the $C_{M_z}\sim\mathcal{R}^{-1}$ decrease (see dashed line in figure \ref{c_m}), and the onset of WVF to make $C_{M_z}$ decrease again with $\mathcal{R}$. In TVF, the drag increases much faster than what is expected from laminar assumption, in WVF, it increases faster (but not so much): TVF is very efficient in conveying momentum radially, WVF not that much, since part of the energy is involved in axial motion. A drag measurement experiment was conducted for a Newtonian fluid (water) and is reported figure \ref{c_m}, for a turbulent state ($\mathcal{R} \gg 1800$). The slope of $C_{M_z}$ is compared with Von Kármán gap theory \citep{Aydin1991Plane-CouetteWalls}. The results show that our $C_{M_z}$ follows the $1 / \sqrt{C_{M_z}} = 3 \ln ( \mathcal{R} \sqrt{C_{M_z})} -2.7 $ corresponding to $\sim \mathcal{R}^{-1/3}$ against the work of \cite{Greidanus2015TurbulentRotation}.

Back to the viscoelastic case, for the higher El fluids, the onset of EIT leads to a sharp increase in $C_{M_z}$ and $\mathcal{N}$. Yet, after this sharp onset, $\mathcal{N}$ tends to a constant value (see figure \ref{alpha_beta_Nu_Ta}), and the friction coefficient asymptotically moves back to its "laminar" evolution, approximately ($\sim \mathcal{R}^{-1}$), as shown in figure \ref{c_m}. This suggests that the additional dissipation induced by the polymer chains and additional radial momentum transport is not Reynolds-dependent. The $\mathcal{R}$-dependency on friction can be scaled by that of the laminar case, with no significant $\mathcal{R}$-dependent contribution of the viscoelastic secondary flows.
This supports the very recent DNS (Direct Numerical Simulation) of \cite{Lopez2022VortexFlow} suggesting that elasto-inertial TCF structures are not efficient in radial momentum convection, which is why they tend to merge and split \citep{Lacassagne2020VortexFlow} or create defects \citep{Latrache2016Defect-mediatedFlow}, thus helping transition to chaos. It is yet worth noting that the jump, i.e the additional dissipative contribution of the polymer, is itself El-dependent: the jump is higher when the elastic number increases. Between the jump and the asymptotic high El region, there is a transitional behaviour which can be understood as the establishment period of EIT. This establishment period displays a different slope, about $\mathcal{R}^{-2/3}$ for the 50 ppm case as illustrated on figure \ref{c_m}. For 25 ppm, the constant $C_{M_z}$ on a narrow range shows the existence of intermediate regimes visually closer to TVF (RSW for Rotating Spiral Waves, see \cite{Lacassagne2020VortexFlow}), and apparently also in terms of torque dynamics (see \cite{Moazzen2022TorqueSuspensions}). Results also show that EIT is different from inertial turbulence modified by polymer for which drag is reduced. It would be interesting to see what happens when $\mathcal{R}$ tends to values for which inertial turbulence is expected, as done in \cite{Samanta2013Elasto-inertialTurbulence} and confront with drag reduction theory.

\subsection{Spatio-temporal analysis: frequency maps and spatial PSD }
\begin{figure}[ht!]
    \centering
    \includegraphics[width=1\textwidth]{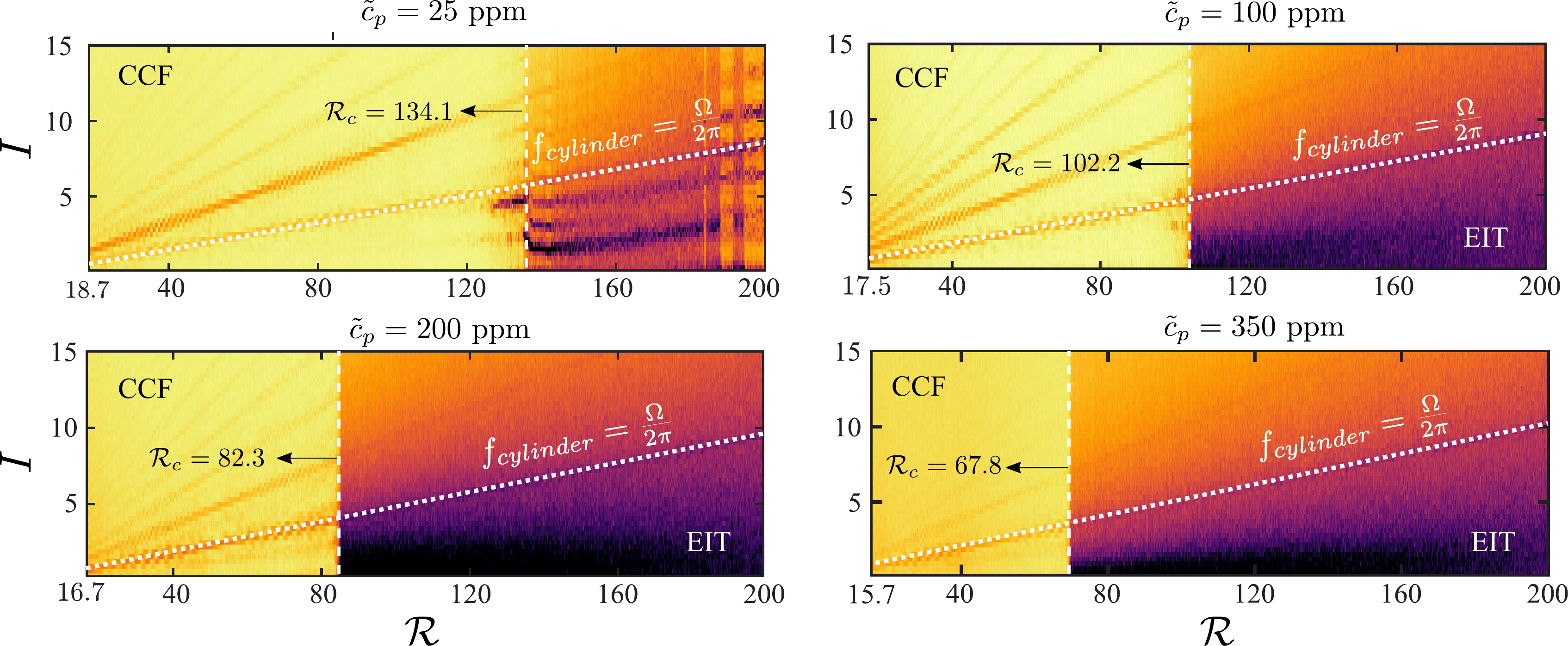}
    \caption{Frequency maps computed as \cite{Moazzen2022TorqueSuspensions} or \cite{Lacassagne2020VortexFlow}, from the reflected light intensity signal, for all vertically stacked $\mathcal{R}$. The color scale represents the FFT intensity from weak (yellow/light) to strong (purple/dark) energy content. Instantaneous change in intensity indicates the transition from CCF to EIT and the oblique dashed lines indicate the rotation frequency of the inner cylinder. For 25 ppm the first instability is RSW which results in the appearance of discrete peaks that are no longer seen in EIT.}
    \label{fig:frequencymap_viscoelastic}
\end{figure}

Following the protocol detailed in \citep{Moazzen2022TorqueSuspensions}, frequency maps for $\tilde{c}_p= 25, 100, 200, 350$ are shown in figure \ref{fig:frequencymap_viscoelastic}. Those plots shows the temporal FFT spectra of the reflected light intensity signal, for all $\mathcal{R}$ stacked vertically, the colorbar representing the intensity of the spectra (arbitrary units). Frequency maps display no particular spectral signature in the CCF domain, before $\mathcal{R}_c$ is reached, other than that of the inner cylinder rotation frequency $f_{cylinder}$ and its harmonics \citep{Cagney2020Taylor-CouetteRheology,Lacassagne2020VortexFlow}. Such ridges are still slightly visible for $\mathcal{R}>\mathcal{R}_c$, but the overall spectral signature changes, with a broadband distribution of energy from large to small temporal scales, a signature of the chaotic behaviour characteristic of EIT \citep{Lacassagne2020VortexFlow,Groisman1996Couette-TaylorSolution,Dutcher2013EffectsFlows}.

\begin{figure}[H] 
    \centering
    \includegraphics[width=1\textwidth]{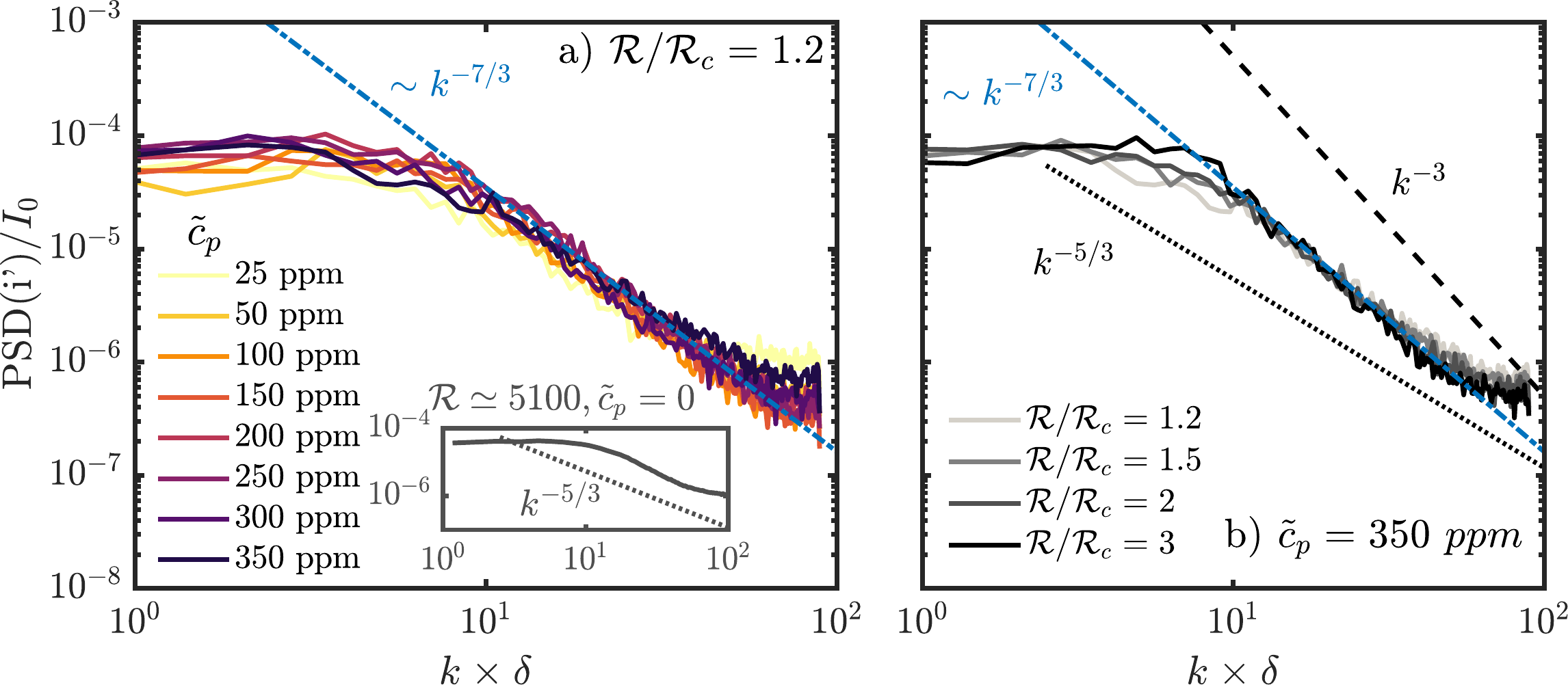}
    \caption{Spatio-temporal dynamics: a) Spatial PSD spectra for all polymer concentrations at $\mathcal{R}/\mathcal{R}_c=1.2$, with an inset showing  typical spectrum for inertial turbulence in water. b) Spatial PSD spectra for $\tilde{c}_p=350~ppm$ at various $\mathcal{R}$ dotted and dashed black lines in b) denote $k^{-5/3}$ and $k^{-3}$ slopes \cite{Yamani2021SpectralTurbulence,Fouxon2003SpectraSolutions}, respectively. The dotted blue line in a) and b) is a -7/3 slope fitting the experimental data.}
    \label{fig:SpatialPSD}
\end{figure}

Further analysis of the spectral behaviour of EIT can be made by computing spatial power spectral density (PSD) of the intensity signal at constant $\mathcal{R}_c$ and $\text{El}$. This is achieved by extracting sets of n vertical lines $I^n(z)$ at constant $\mathcal{R}$ on flow maps (e.g from figure \ref{Nu_st}), subtracting the average intensity profile $\left<I\right>(z)=\frac{1}{n}\sum_{j=1}^n I^j(z)$ in order to define the intensity fluctuation $(i')^n(z)=I^n(z)-\left<I\right>(z)$, computing PSD($(i')^n(z))$ and averaging PSDs on $n$. Values are finally scaled by the spatial (over $z$) average of $\left<I\right>(z)$ called $I_0$. One typically uses $n=50$, a number for which the convergence of spectra was deemed sufficient for the analysis that follows. PSD spectra for all polymer concentrations at $\mathcal{R}/\mathcal{R}_c\simeq 1.2$ are reported in figure  \ref{fig:SpatialPSD} a), and spectra for the 350~ppm case at various $\mathcal{R}$ values in figure \ref{fig:SpatialPSD} b). An additional experiment was performed with water, in order to reach $\mathcal{R}\simeq 5100$ and produce a flow map and PSD in the inertial turbulence regime (inset of figure \ref{fig:SpatialPSD} a). 
For that curve, one would expect to capture a -5/3 slope if i) turbulence is sufficiently developed and ii) the intensity fluctuation signal is in some way representative of the radial velocity fluctuations, as suggested by \cite{Abcha2008QualitativeSystem}. It appears from the inset of \ref{fig:SpatialPSD} a) that the inertial turbulence curve roughly follows the -5/3 trend at least at intermediate scales, which seems to validate both arguments. When considering spectra for EIT, the -5/3 slope is not expected in low $\mathcal{R}$ cases where inertial turbulence would not have been present. Different scaling exponents of EIT in various flow configurations (channel flows, TC flows...) can be found in the literature, in a range from -14/3 to -3, always steeper than the inertial case. The -3 slope has recently been put forward by \cite{Yamani2021SpectralTurbulence} in their study of viscoelastic polymer jets as a universal spectral behaviour of EIT, after having been theoretically predicted by \cite{Fouxon2003SpectraSolutions}. Recent numerical simulations by \cite{Lopez2022VortexFlow} have been able to retrieve this scaling in elasto-inertial TCF. This reference slope is also represented on figure \ref{fig:SpatialPSD} b).  

\raggedbottom

It here appears that all curves for EIT fall short of -3 trends, but still display a somehow universal slope with respect to polymer concentration (figure \ref{fig:SpatialPSD} a). An increase in Reynolds number seems to consolidate this slope by increasing the $k$ span over which it applies (figure \ref{fig:SpatialPSD} b). The slope value is here around -7/3 (figure \ref{fig:SpatialPSD} a and b). It is worth mentioning that the visualisation method probes the flow from the outside and not in the bulk, and may be subject to boundary layer effects on the outer cylinder. So the value of the slope itself must be interpreted with care. Figure \ref{fig:SpatialPSD} yet confirms key findings \citep{Yamani2021SpectralTurbulence} namely that of an apparent universal spectral slope of EIT, steeper than inertial turbulence. Additionally, our results suggest that EIT is intrinsically a combination of elasticity and inertia, as $\mathcal{R}$ helps develop the slope when increased.

\section{Conclusion}
In this work, new characterisations of the dynamics of EIT in Taylor-Couette flow of polymer solutions were presented. Combining flow visualisation and torque measurements allowed to detect CCF-EIT transition and to describe key dynamic features of EIT in TCF. In particular the scaling of $\mathcal{N}$ with Ta and its dependency on fluid elasticity has been discussed. Two sub-domains of EIT were reported: a transitional one for which energy dissipation is still dominated by inertia and a fully developed one for which elastic energy transfer become dominant. Spectral analysis support the idea of a chaotic developed EIT state for which PSD would exhibit a universal slope. Developed EIT displays an asymptotic "laminar-like" scaling for the friction coefficient: the wall friction is directly correlated to the base azimuthal flow and secondary flow structures do not play a role in wall friction as their energy is dissipated elastically. This requires a sufficient level of both elasticity and inertia, and is expected to be of great interest in the aim of achieving efficient mixing at low drag and low $\mathcal{R}$.

\enlargethispage{20pt}

%
%\ethics{Insert ethics statement here if applicable.}

%\dataccess{Insert details of how to access any supporting data here.}

\aucontribute{MM performed, analysed, interpreted the experimental work reported in this paper, and drafted the initial manuscript. TL and AB equally contributed to the design of the experiments (together with MM), conceptualisation of the project, guidance, and supervision of the work (together with VT). AB and VT secured funding. All authors contributed to the writing, reading, editing and approved the manuscript.}  

\competing{The author(s) declare that they have no competing interests.}

\funding{Financial support from Region Hauts de France and Institut Mines Télécom is gratefully acknowledged.}

%\ack{Insert acknowledgment text here.}

%\disclaimer{Insert disclaimer text here if applicable.}

%%%%%%%%%% Insert bibliography here %%%%%%%%%%%%%%
\bibliographystyle{ieeetr}
\bibliography{references}

\end{document}